\documentclass[aps,prl,twocolumn,showpacs,amsmath,floatfix,superscriptaddress]{revtex4-1}

\usepackage{graphicx}
\usepackage{amsmath}
\usepackage{braket}
\usepackage{siunitx}
\usepackage{bbold}

\begin{document}

\title{Nonanalyticity, Valley Quantum Phases, and Light-like Exciton Dispersion in Monolayer TMDs: Theory and First-Principles Calculations}

\author{Diana Y. Qiu}
\affiliation{Department of Physics, University of California at
Berkeley, California 94720}
\affiliation{Materials Sciences Division, Lawrence Berkeley
National Laboratory, Berkeley, California 94720}

\author{Ting Cao}
\affiliation{Department of Physics, University of California at
Berkeley, California 94720}
\affiliation{Materials Sciences Division, Lawrence Berkeley
National Laboratory, Berkeley, California 94720}

\author{Steven G. Louie}
\email[Email: ]{sglouie@berkeley.edu}
\affiliation{Department of Physics, University of California at
Berkeley, California 94720}
\affiliation{Materials Sciences Division, Lawrence Berkeley
National Laboratory, Berkeley, California 94720}

\begin{abstract}
Exciton dispersion as a function of center-of-mass momentum, \textbf{Q}, is essential to the understanding of exciton dynamics. We use the ab-initio GW-Bethe Salpeter equation method to calculate the dispersion of excitons in monolayer MoS$_2$ and find a nonanalytic light-like dispersion. This behavior arises from an unusual $|Q|$ term in both the intra- and inter-valley exchange of the electron-hole interaction, which concurrently gives rise to a valley quantum phase of winding number two. A simple effective Hamiltonian to $Q^2$ order with analytic solutions is derived to describe quantitatively these behaviors. \end{abstract}

\pacs{73.22.-f, 71.35.-y, 78.67.-n}

\maketitle

An exciton is a neutral excitation of an interacting electron system, consisting of a bound excited electron and hole pair with energy depending on its center-of-mass momentum \textbf{Q}. In crystals, the exciton dispersion relation (energy vs. \textbf{Q}) forms an exciton band structure. In optical absorption, the momentum of a photon is converted to that of an exciton, resulting in a small \textbf{Q} exciton for visible light. Recent model Hamiltonian calculations predict that the dispersion of the lowest energy optically active exciton bands in monolayer MoS$_2$ form a Dirac cone (as in graphene) due to the exchange interaction coupling electron-hole states in different valleys (the intervalley exchange)~\cite{yu14}. Another study based on a tight-binding formulation of the BSE approach suggested also linear dispersion for the lowest energy excitons near $\textbf{Q}=0$, although the dispersion relation (on a $45\times45$ finite-Q grid) was not resolved at the lengthscale of the momentum of light~\cite{wu15}. The effects of the interplay of inter- and intra-valley exchange and local fields are not explicitly investigated in these model calculations, which also miss terms in both $Q$ and $Q^2$ orders.

In this work, we calculate the exciton dispersion of MoS$_2$, a prototypical transition metal dichalcogenide (TMD), from first principles, using the ab-initio GW-Bethe Salpeter equation (GW-BSE) method~\cite{hybertsen86,rohlfing00}, and find a highly unusual low-energy dispersion consisting of a nonanalytic v-shaped upper band that is degenerate with a parabolic lower band at $\textbf{Q}=0$, consistent qualitatively with previous tight-binding results~\cite{wu15}. We show that the physical origin of this highly nonanalytic behavior (the $|Q|$-dependence) comes from $|Q|$-dependent terms in both the intravalley and intervalley exchange interaction, which arise from the unique electronic structure and the quasi-2D nature of the Coulomb interaction in atomically thin TMDs. Local-field effects introduce additional interaction terms and are responsible for the splitting of optically bright and dark excitons. Moreover, the theory gives a valley quantum phase of winding number two (or chirality two), which we show should manifest in optical experiments as a phase difference between the longitudinal and transverse response. A similar winding number is found in Ref.~\cite{yu14}, although in this model Hamiltonian study, intravalley exchange is neglected resulting in a Dirac cone dispersion.

Transition metal dichalcogenides are layered, weakly-coupled materials. In the 2H monolayer form, TMDs are direct band gap semiconductors with the gap at the $K$ and $K$' points of the 2D hexagonal Brillouin zone\cite{mak10,splendiani10}. Due to a lack of inversion symmetry and presence of three-fold rotational symmetry, a valley selective response emerges: the $K$ and $K$' valley couples respectively only to left or right circularly polarized light~\cite{cao12,xiao12,zeng12,mak12}. Moreover, as a result of confinement and reduced screening, excitons have very large binding energies~\cite{lambrecht12,ashwin12,komsa12,shi13,qiu13,bernardi13,conley13,espejo13,huser13,soklaski14,ugeda14,chernikov14,ye14,klots14,zhu14,zhang14}.

\begin{figure}
    \includegraphics[width=246.0pt]{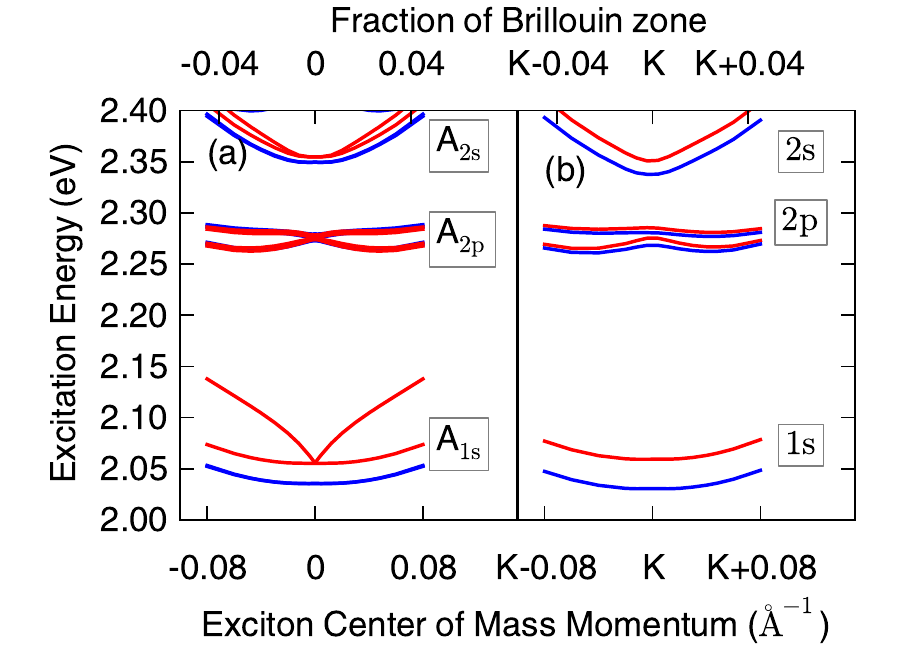}
    \caption{\label{fig:optbs}
    (Color online) Exciton dispersion of monolayer MoS$_2$ near (a) \textbf{Q}=0 and (b) \textbf{Q}=K along the $K-\Gamma-K'$ direction. Red(blue) lines indicate states arising from like(unlike)-spin transitions. The label A refers to states involving transitions from the highest valence band at K/K'. "B"-states involving transitions from the second highest valence band are similar and shown in the Supplemental Materials.}
\end{figure}

The quasiparticle (QP) bandstructure of monolayer MoS$_2$ calculated within the GW approximation\cite{hybertsen86,qiu13} (see Supplemental Materials) is shown in Fig.~\ref{fig:qpbs}. The direct gap of 2.67~eV is at the $K$ and $K$' points. Spin-orbit coupling splits the valence band edge by 147~meV and the conduction band edge by 3~meV. Over an extended range in the $K$ and $K$' valleys, $s_z$, the spin of the electron along the direction perpendicular to the layer, is a good quantum number \cite{xiao12}. Thus, the concept of singlet and triplet exciton states is no longer well-defined. Instead, the electron-hole Hamiltonian or BSE matrix can be decoupled into transitions between  bands of like-spin (Fig.~\ref{fig:qpbs}(b-d)) that are optically allowed and transitions between bands of unlike spin that are optically forbidden. We calculate the exciton dispersion of both the like-spin transition states and the unlike-spin transition states within the BSE formalism following Ref.\cite{rohlfing00}. The resulting exciton dispersion near \textbf{Q}=0 and \textbf{Q}=K is shown in Fig.~\ref{fig:optbs}. Fig.~\ref{fig:optbs} only presents exciton states involving predominantly transitions from the topmost valence band (the $v\uparrow K$ and $v\downarrow K'$ bands in Fig.~\ref{fig:qpbs}), which is the so-called A series in the literature. Similar results are obtained for the spin-orbit split B series (see Supplemental Materials).

\begin{figure*}
    \includegraphics[width=492.0pt]{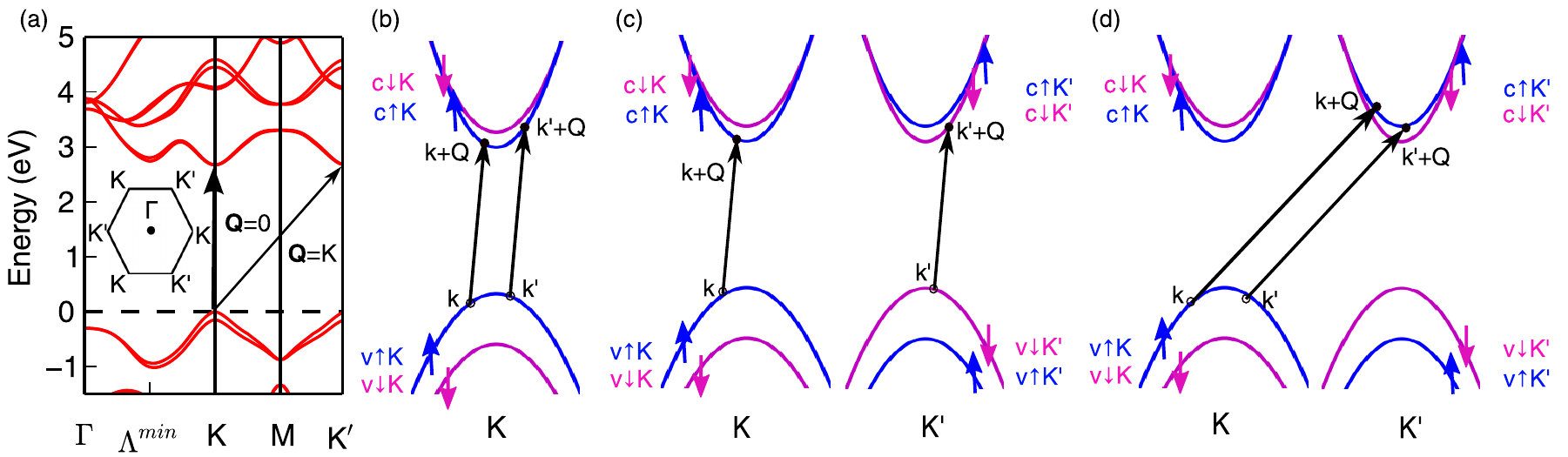}
    \caption{\label{fig:qpbs}
    (Color online) (a) QP bandstructure of monolayer MoS$_2$. Lowest energy transitions corresponding to electron-hole excitations with momentum transfer of $\textbf{Q}=0$ and $\textbf{Q}=K$ are shown with the labeled arrows. (b-c) Schematics of interactions between two electron-hole pairs with momentum near $\textbf{Q}=0$, corresponding to BSE matrix elements $\langle vc\textbf{k}\textbf{Q}|H^{BSE}|v'c'\textbf{k}'\textbf{Q}\rangle$ for: (b) two like-spin transitions within one valley and (c) two like-spin transitions from different valleys. (d) Schematic of two electron-hole pairs with momentum near $\textbf{Q}=K$ giving rise to BSE matrix elements for like-spin transitions.
    }
\end{figure*}

In Fig.~\ref{fig:optbs}, near $\textbf{Q}=0$, the lowest energy exciton complex shown is the "1s"-like states of the A series excitons, which has a binding energy of $0.63$~eV for the like-spin transition states and $0.65$~eV for the unlike-spin transition states. At $\textbf{Q}=0$, the two-fold degeneracy of the exciton states due to $K$ and $K$' valley degeneracy is protected by the time-reversal symmetry. Away from $\textbf{Q}=0$, for optically active like-spin transition states, the degenerate bands split as $|Q|$ increases. Consistent with previous model calculations\cite{yu14,wu15,yu14_prb,glazov14}, this splitting of the like-spin transition states is due to intervalley exchange. The exciton bands for unlike-spin states, which have zero exchange interaction, remain doubly degenerate (the blue lines in Fig.~\ref{fig:optbs}). However, contrary to previous model predictions~\cite{yu14}, our ab-initio results do not find a Dirac cone near $\textbf{Q}=0$ in the dispersion of the like-spin transition "1s" bands. In fact, we find a "v-shaped" nonanalytic upper band and a parabolic lower band, with both bands increasing monotonically with $|Q|$ down to the smallest sampled $|Q|$ of $4\times10^{-3}\AA^{-1}$, which corresponds to 0.3\% of the distance from $\Gamma$ to $K$ in the Brillouin zone.


We also calculate the dispersion of the intervalley excitons (Fig.~\ref{fig:optbs}(b))---i.e., an exciton with the electron and hole in different valleys (with $\textbf{Q}=K$)---and find that the unlike-spin transitions state is slightly lower in energy than the similar exciton at $\textbf{Q}=0$ as a consequence of the spin-orbit splitting of the conduction bands and the screened Coulomb interaction (see Supplemental Materials).

\begin{figure}
    \includegraphics[width=246.0pt]{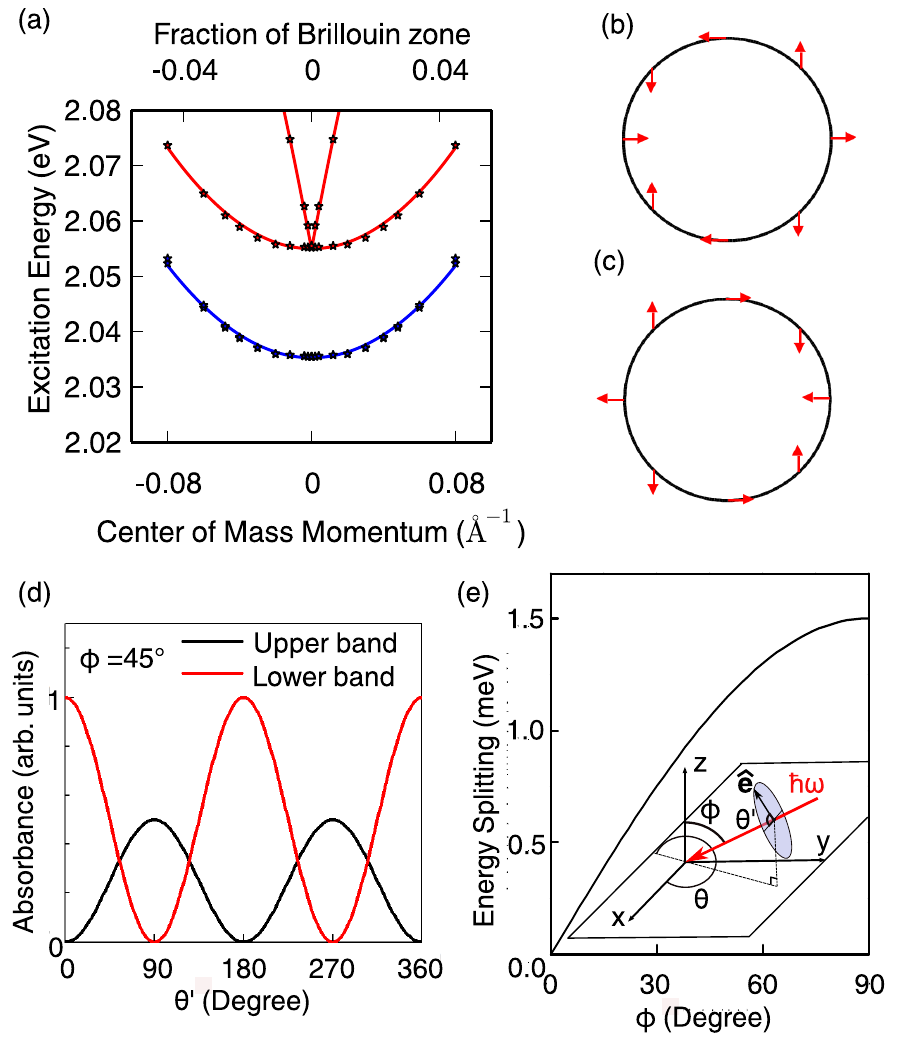}
    \caption{\label{fig:1s}
    (Color online) (a) Close-up of dispersion of $A_{1s}$ near \textbf{Q}=0. Ab-initio values are stars. Fit to effective Hamiltonian (Eq.~\ref{eqn:hexch}) are solid lines. Red(blue) lines indicate states arising from like(unlike)-spin transitions. (b) Valley pseudospin texture of the (upper) nonanalytic like-spin transition band around \textbf{Q}=0 for states of fixed energy in \textbf{Q}-space. (c) Valley pseudospin texture of the (lower) parabolic like-spin transition band. (d) Optical absorbance of linearly polarized light at fixed incidence as the polarization vector $\hat{\textbf{e}}$ is rotated over $360^\circ$.  The angle of the polarization vector, $\theta'$, is defined with respect to the vector formed by the intersection of the polarization plane(blue) and the xy-plane. Red(black) indicates the absorbance of states arising from the lower (upper) like-spin band. (e) Energy difference between the upper and lower like-spin bands that is probed as $\phi$, the angle between wavevector of light and the z-axis, is changed, for light of $\hbar\omega\approx 2$~eV. The inset shows how $\theta$, $\phi$, and $\theta'$ are defined.}
\end{figure}

We now provide the physical origin of the nonanalytic behavior of the dispersion seen in Fig.~\ref{fig:optbs}(a) (closeup in Fig.~\ref{fig:1s}a). The exciton dispersion is obtained from first principles by solving the Bethe-Salpeter equation (BSE)~\cite{rohlfing00,jdeslip11BGW} for electron-hole pair excitations with finite center-of-mass momentum \textbf{Q}
\begin{flalign}
\begin{split}
\label{eqn:BSE}
&(E_{c\textbf{k}+\textbf{Q}}-E_{v\textbf{k}})A^{S}_{vc\textbf{kQ}} + \\ &\sum_{v'c'\textbf{k}'}\langle vc\textbf{kQ}|K^{eh}|v'c'\textbf{k}'\textbf{Q} \rangle A^S_{v'c'\textbf{k}'\textbf{Q}} = \Omega^S_{\textbf{Q}}A^{S}_{vc\textbf{kQ}}.
\end{split}
\end{flalign}
Here, $S$ indexes the exciton states; $A^{S}_{vc\textbf{kQ}}$ is the amplitude of the free electron-hole pair consisting of an electron in $|c\textbf{k}+\textbf{Q}\rangle$ and one missing from $|v\textbf{k}\rangle$; $\Omega^S_\textbf{Q}$ is the exciton excitation energy; $E_{c\textbf{k}+\textbf{Q}}$ and  $E_{v\textbf{k}}$ are the quasiparticle energies, and $K^{eh}$ is the electron-hole interaction kernel. The kernel consists of a direct term and an exchange term\cite{rohlfing00}: $\langle vc\textbf{kQ}|K^{eh}|v'c'\textbf{k}'\textbf{Q} \rangle = \langle vc\textbf{kQ}|K^{d}+K^{x}|v'c'\textbf{k}'\textbf{Q} \rangle $.
The exchange term is
\begin{flalign}
\label{eqn:KxG}
\begin{split}
\langle vc\textbf{kQ}|&K^{x}|v'c'\textbf{k}'\textbf{Q} \rangle = \\
&\sum_\textbf{G}
M_{cv}(\textbf{k},\textbf{Q},\textbf{G})
v(\textbf{Q}+\textbf{G})
M_{c'v'}^*(\textbf{k}',\textbf{Q},\textbf{G})
\mathrm{,}
\end{split}
\end{flalign}
where $\textbf{G}$ are reciprocal lattice vectors; $v$ is the bare Coulomb interaction, and $M$ is defined as $M_{nn'}(\textbf{k},\textbf{Q},\textbf{G})=
\langle n\textbf{k}+\textbf{Q}| e^{i(\textbf{Q}+\textbf{G})\cdot \textbf{r}}|n'\textbf{k}\rangle
$.
The exchange term is only non-zero for like-spin transitions (i.e. excitons with total spin along $z$ equal to zero). The direct term is
\begin{flalign}
\label{eqn:KdG}
\begin{split}
&\langle vc\textbf{kQ}|K^{d}|v'c'\textbf{k}'\textbf{Q} \rangle = \\
&-\sum_{\textbf{GG}'}
M_{cc'}^*(\textbf{k}+\textbf{Q},\textbf{q},\textbf{G})
W_{\textbf{GG}'}(\textbf{q})
M_{vv'}(\textbf{k},\textbf{q},\textbf{G})
\mathrm{,}
\end{split}
\end{flalign}
where $\textbf{q}=\textbf{k}-\textbf{k}'$ and $W$ is the screened Coulomb interaction.

The solutions of the BSE matrix are exciton states $|S_\textbf{Q}\rangle=\sum_{vc\textbf{k}}A^{S}_{vc\textbf{k}\textbf{Q}}|vc\textbf{k}\textbf{Q}\rangle$.
In the subspace spanned by the two lowest energy states, at exactly $\textbf{Q}=0$, the exchange matrix elements are also diagonal.
At $\textbf{Q}=0$, the exchange term is the diagonal constant
\begin{flalign}
\begin{split}
\label{eqn:C}
&C\equiv \langle S_\textbf{0}|K^x|S_\textbf{0}'\rangle = 
\delta_{SS'} 2\pi e^2\sum_{\textbf{G}\neq 0}  
\sum_{vc\textbf{k}v'c'\textbf{k}'} \\ &A^{S*}_{vc\textbf{k}\textbf{0}}A^{S'}_{v'c'\textbf{k}'\textbf{0}}\frac{\langle u_{c\textbf{k}}|e^{i\textbf{G}\cdot \textbf{r}}|u_{v\textbf{k}}\rangle(\langle u_{c'\textbf{k}'}|e^{i\textbf{G}\cdot \textbf{r}}|u_{v'\textbf{k}'}\rangle)^*}{|G|}
\mathrm{,}
\end{split}
\end{flalign}
where for small \textbf{Q}'s in $2D$, $v(\textbf{Q}+\textbf{G})=\frac{2\pi e^2}{|\textbf{Q}+\textbf{G}|}\approx\frac{2\pi e^2}{|\textbf{G}|}$. This constant term is $20$~meV in our ab-initio calculation. It is purely a local-field effect and is responsible for the splitting between the like-spin(bright) and unlike-spin(dark) states at $\textbf{Q}=0$.
 
We now derive an effective Hamiltonian ($H^{BSE}$) to describe the main physics for the 1s complex in a basis of "excitonic" functions from the individual valleys in the tight-binding limit, given in Eq.~\ref{eqn:basis} below. At $\textbf{Q}=0$, both the like and unlike-spin excitonic levels are doubly degenerate, with the amplitude $A^{S}_{vc\textbf{k0}}$ of one state ($|S^K_\textbf{0}\rangle$) confined to the $K$ valley and that of the other ($|S^{K'}_\textbf{0}\rangle$) confined to the $K$' valley. Near $\textbf{Q}=0$, it is sufficient to use the following basis functions (which are of the Bloch form of excitons from a specific valley in the tight-binding limit) to expand the true exciton state $|S_\textbf{Q}\rangle$:
\begin{flalign}
\label{eqn:basis}
\begin{split}
|S^K_\textbf{Q}\rangle \approx 
 |e^{i\textbf{Q}\cdot \textbf{R}}S^K_\textbf{0}\rangle
 \equiv \left(\begin{matrix} 1 \\ 0 \end{matrix}\right)_{\textbf{Q}}  ,
 |S^{K'}_\textbf{Q}\rangle \approx 
|e^{i\textbf{Q}\cdot \textbf{R}}S^{K'}_\textbf{0}\rangle \equiv \left(\begin{matrix} 0 \\ 1 \end{matrix}\right)_\textbf{Q}
\mathrm{,}
\end{split}
\end{flalign}
where $\textbf{R}=\frac{\textbf{r}_e+\textbf{r}_h}{2}$,
$(\begin{smallmatrix} 1 \\ 0 \end{smallmatrix})_\textbf{Q}$ and $(\begin{smallmatrix} 0 \\ 1 \end{smallmatrix})_\textbf{Q}$ are pseudospinors denoting respectively basis functions on $K$ and $K'$ valleys, and $|e^{i\textbf{Q}\cdot \textbf{R}} S^K_\textbf{0}\rangle=\sum_{vck} A^S_{vc\textbf{k0}}|vc\textbf{kQ}\rangle$.

In this basis, $H^{BSE}$ is a $2\times2$ matrix, and the intravalley exchange term (matrix element between basis functions in the same valley) is
\begin{flalign}
\label{eqn:intra}
\begin{split}
\langle S_\textbf{Q}^{K}|K^x|S_\textbf{Q}^{K}\rangle
=\sum_{vc\textbf{k}v'c'\textbf{k}'} A^{S^{K}*}_{vc\textbf{k}\textbf{0}}A^{S^{K}}_{v'c'\textbf{k}'\textbf{0}} \langle vc\textbf{kQ}|K^x|v'c'\textbf{k}'\textbf{Q}\rangle .
\end{split}
\end{flalign}
Using a $\textbf{Q}\cdot \textbf{p}$ expansion of the QP states in the $M$ matrix elements (Eq.~\ref{eqn:KxG}) to second order in $\textbf{Q}\cdot \textbf{p}$, the intravalley exchange term (Eq.~\ref{eqn:intra}) becomes
\begin{flalign}
\begin{split}
\label{eqn:Kx_kdotp}
&\langle S_\textbf{Q}^{K}|K^x|S_\textbf{Q}^{K}\rangle  = \langle S_\textbf{Q}^{K'}|K^x|S_\textbf{Q}^{K'}\rangle =
\sum_{vc\textbf{k}v'c'\textbf{k}'} A^{S^{K}*}_{vc\textbf{k}\textbf{0}}A^{S^K}_{v'c'\textbf{k}'\textbf{0}} \\
& \{ \textbf{Q}\cdot a(\langle u_{v\textbf{k}}|\textbf{p}|u_{c\textbf{k}}\rangle)^*
\textbf{Q}\cdot a'\langle u_{v'\textbf{k}'}|\textbf{p}|u_{c'\textbf{k}'}\rangle v(\textbf{Q}) \\
& + \sum_{\textbf{G}\neq 0} v(\textbf{Q}+\textbf{G})
[c + O(\textbf{Q}\cdot \textbf{p}) + O((\textbf{Q}\cdot \textbf{p})^2)]
\}
.
\end{split}
\end{flalign}
Here, we have separated contributions from the $\textbf{G}=0$ and $\textbf{G}\neq 0$ Fourier components (Higher order terms are described explicitly in the Supplemental Materials). We note that $a$, $a'$, and $c$ are factors that depend on the QP states but are independent of \textbf{Q}
\footnote{\unexpanded{$a=\frac{\hbar}{m}(E_{c\textbf{k}}-E_{v\textbf{k}})^{-1}$. $a'$ is equal to $a$ with $(\textbf{k}\leftrightarrow\textbf{k}'$. $c=\langle u_{c\textbf{k}}|e^{i\textbf{G}\cdot \textbf{r}}|u_{v\textbf{k}}\rangle(\langle u_{c'\textbf{k}'}|e^{i\textbf{G}\cdot \textbf{r}}|u_{v'\textbf{k}'}\rangle)^*$.}}.
In Eq.~\ref{eqn:Kx_kdotp}, the $\textbf{G}=0$ term, has a nonanalytic dependence on  $|Q|$
\begin{flalign}
\begin{split}
|\textbf{Q}\cdot\sum_{vc\textbf{k}} a A^{S^K}_{vc\textbf{k}\textbf{0}} \langle u_{v\textbf{k}}|\textbf{p}|u_{c\textbf{k}}\rangle|^2 v(\textbf{Q})
 \propto
A|Q|
\mathrm{.}
\end{split}
\end{flalign}
Here, $A$ is a proportionality constant, and we have made use of the fact that for small $\textbf{Q}$ in $2D$, $v(\textbf{Q})\approx\frac{2\pi e^2}{|Q|}$, and the fact that $\langle 0|\textbf{p}|S^K_\textbf{0}\rangle\propto \hat{x}+i\hat{y}$ due to the $C_3$ symmetry\cite{cao12}, which eliminates the dependence on the orientation of \textbf{Q}. Up to order $Q^2$, the intravalley exchange (Eq.~\ref{eqn:intra})  has the form
\begin{equation}
\label{intra_simp}
\langle S_\textbf{Q}^{K}|K^x|S_\textbf{Q}^{K}\rangle =
\langle S_\textbf{Q}^{K'}|K^x|S_\textbf{Q}^{K'}\rangle =
C + A|Q| +\beta Q^2
.
\end{equation}
$C$ is the splitting of bright and dark states given in Eq.~\ref{eqn:C}, and $\beta$ is a proportionality constant that is a real number, which arises from the local fields. Likewise, the intervalley exchange may be shown to be
\begin{equation}
\label{eqn:inter_simp}
\langle S_\textbf{Q}^{K}|K^x|S_\textbf{Q}^{K'}\rangle =
(\langle S_\textbf{Q}^{K'}|K^x|S_\textbf{Q}^{K}\rangle)^* = 
A|Q|e^{-i2\theta} + \beta' Q^2
\mathrm{,}
\end{equation}
where $\theta$ is the angle of \textbf{Q} defined with respect to the x-axis and $\beta'$ is a complex number.

After performing a similar analysis of the direct term, we find an effective Hamiltonian
\begin{widetext}
\begin{equation}
\label{eqn:hexch}
H^{BSE}(\textbf{Q})
=
\Omega_\textbf{0}\mathbb{1}
+ 
A[\mathbb{1}+\cos(2\theta)\sigma_x+\sin(2\theta)\sigma_y]|Q|
+
[(\frac{\hbar^2}{2M}+\alpha + \beta )\mathbb{1} +\beta'(\sigma_x + \sigma_y)]Q^2
\mathrm{,}
\end{equation}
\end{widetext}
where $\Omega_\textbf{0}$ is the excitation energy of the exciton with $\textbf{Q}=0$; $M=m_e+m_h$ is the QP band mass of the free electron-hole pair at $K$ or $K'$; $\alpha$ is a constant from the order of $Q^2$ contribution from the direct term; $\beta$ and $\beta'$ are constants from the order of $Q^2$ contribution from the intravalley and intervalley exchange respectively; $\sigma_x$ and $\sigma_y$ are the Pauli matrices (see Supplemental Materials for detailed derivation). An effective Hamiltonian may be obtained for the other exciton complexes through similar analyses.

Diagonalizing the effective Hamiltonian (Eq.~\ref{eqn:hexch}) gives us two solutions: one with a parabolic dispersion
\begin{equation}
\label{eqn:omega1}
\Omega_{-}(\textbf{Q})
=\Omega_\textbf{0}+\left(\frac{\hbar^2}{2M}+\alpha + \beta - |\beta'|\right)Q^2
\mathrm{,}
\end{equation}
and the other with nonanalytic dispersion
\begin{equation}
\label{eqn:omega2}
\Omega_{+}(\textbf{Q})=\Omega_\textbf{0}+2A|Q|+\left(\frac{\hbar^2}{2M}+\alpha + \beta +  |\beta'| \right)Q^2 
\end{equation}
Hence, it is the combination of intervalley and intravalley exchange that results in a lower band with a parabolic dispersion (Eq.\ref{eqn:omega1}) and an upper band with a nonanalytic dependence on $|Q|$ (Eq.~\ref{eqn:omega2}), as seen in our ab-initio calculation (Fig.~\ref{fig:1s}a). We emphasize that this $|Q|$-dependence at small \textbf{Q} is a consequence of the 2D Coulomb interaction and the opposite angular polarization of the electronic states in the $K$ and $K'$ valleys, which removes the dependence on the direction of \textbf{Q}.

We fit Eqs. ~\ref{eqn:omega1} and ~\ref{eqn:omega2} to our ab-initio calculation (Fig.~\ref{fig:1s}(a)) to obtain the values of the proportionality constants, $A$, $\alpha$, $\beta$, and $\beta'$. 
We find $A=0.9$~eV$\cdot\AA$. The dominant $Q^2$ term comes from the QP effective mass $\frac{\hbar^2Q^2}{2M}$. $M$ for monolayer MoS$_2$ is roughly $1.1m_0$. From the dispersion of the unlike-spin states, we find that $\alpha=-0.9$~eV$\cdot\AA^2$, and the effective mass of the unlike-spin transition(dark), 1s exciton is roughly $M^*=1.5m_0$. The effective mass of the parabolically-dispersing like-spin(bright), 1s exciton is $M^*=1.4m_0$, indicating that $\beta-|\beta'|$, the difference between quadratic terms in the intervalley and intravalley exchange, is small, only about $0.2$~eV$\cdot\AA^2$, while the magnitude of $\beta$ is about $4$~eV$\cdot\AA^2$. Our results demonstrate an enhancement of about 30\% in the center-of-mass effective mass $M^*$ of the parabolic band excitons due to electron-hole interaction~\cite{mattis84}.

Since the difference between the magnitude of the quadratic terms in the intervalley and intravalley exchange, $\beta-|\beta'|$, is small, we may to good approximation neglect this difference (i.e. we take $\beta'=e^{-i2\theta}\beta$. See Supplemental Materials.). Then, the solution for the parabolic like-spin transition exciton band simplifies to
\begin{equation}
\label{eqn:SQ1}
\frac{1}{\sqrt{2}}(e^{-i\theta}|S^K_\textbf{Q}\rangle - e^{i\theta}|S^{K'}_\textbf{Q}\rangle)
\mathrm{,}
\end{equation}
and the solution for the nonanalytic like-spin transition exciton band simplifies to
\begin{equation}
\frac{1}{\sqrt{2}}(e^{-i\theta}|S^K_\textbf{Q}\rangle + e^{i\theta}|S^{K'}_\textbf{Q}\rangle)
\mathrm{.}
\end{equation}
For both states, the valley pseudospin winding number around \textbf{Q}=0 is 2. The pseudospin texture of these states in a circle in \textbf{Q}-space is shown in Fig.~\ref{fig:1s}(b-c). To study the optical response, we project the momentum and polarization of incident linearly-polarized light into the layer $2D$ plane. The momentum \textbf{Q} of the excited exciton is equivalent to the in-plane component of the wavevector of the light. The in-plane projection of the electric field polarization vector can be decomposed into components perpendicular (transverse) to and parallel (longitudinal) to the momentum transfer \textbf{Q}. We define the angle of the polarization vector with respect to the transverse projection as $\theta'$. The absorbance of the upper and lower bands with respect to $\theta'$ has a phase difference of $90^\circ$, and the intensity of each band peaks twice as the polarization angle is rotated over $360^\circ$ (Fig.~\ref{fig:1s}d and Supplementary Materials). The different pseudospin texture results in the upper band coupling to the longitudinal component of the in-plane projection of the electric field and the lower band coupling to the transverse component of the in-plane electric field (see Supplemental Materials). Since the transverse projection is always in the $2D$ plane, only the intensity of optical absorbance from the upper band changes with the angle of the incident light with respect to the z-axis ($\phi$). Intensity is maximum at normal incidence ($\phi=0$) and minimum at grazing incidence ($\phi\approx90^\circ$). The energy difference between the upper and lower band for different \textbf{Q} on the dispersion curve can also be probed by changing $\phi$. For photons with $\hbar\omega\approx 2$~eV, the largest energy difference is about 1.5~meV when the wavevector of the light is nearly parallel to the plane.

In summary, we have computed the exciton dispersion of MoS$_2$ from first principles and find an unusual dispersion with a parabolic lower band and a v-shaped upper band for the lowest energy like-spin transition exciton complex near $\textbf{Q}=0$. This dispersion is due to the interplay of the intervalley and intravalley exchange, both of which have a $|Q|$-dependent behavior. We have derived a simple effective Hamiltonian and analytic solutions describing this physics and predict that the splitting of the exciton bands can be measured with a linearly polarized optical beam tilted away from normal incidence. We expect any $2D$ semiconductor with excitons with amplitude concentrated in a small portion of the Brillouin zone to exhibit similar nonanalytic exciton dispersion near $\textbf{Q}=0$. We also show that interaction effects increase the exciton mass $M^*$ by $>30\%$. First-principles results for the intervalley excitons with $\textbf{Q}=K$ and other finite \textbf{Q} excitons are also obtained.

\section*{Acknowledgments}

We thank F.H. Jornada, F. Bruneval, J. Deslippe, and T. Rangel for fruitful discussions. This research was supported by the Theory Program (which provided the GW-BSE calculations and simulations) and by the SciDAC Program on Excited State Phenomena in Energy Materials  ( which provided the finite momentum exciton codes) at the Lawrence Berkeley National Lab through the Office of Basic Energy Sciences, U.S. Department of Energy under Contract No. DE-AC02-05CH11231, and by the National Science Foundation under Grant No. DMR10-1006184 (which provided the formulation of the effective Hamiltonian theory). D. Y. Q. acknowledges support from the NSF Graduate Research Fellowship Grant No. DGE 1106400. This research used resources of the National Energy Research Scientific Computing Center, which is supported by the Office of Science of the U.S. Department of Energy, and the Extreme Science and Engineering Discovery Environment (XSEDE), which is supported by National Science Foundation grant number ACI-1053575.


\bibliography{MoS2}
\bibliographystyle{apsrev4-1}

\end{document}